\documentclass[twocolumn,showpacs,preprintnumbers,superscriptaddress,prb,floatfix,aps,10pt]{revtex4-1}

\usepackage{amsmath,amssymb}
\usepackage{xcolor}
\usepackage{graphicx,textcomp}

\DeclareUnicodeCharacter{2009}{\,} 

\begin{document}

\title{Optimization of Quantum-dot Qubit Fabrication via Machine Learning}

\author{Antonio B. Mei}
\email{armei@hrl.com}
\author{Ivan Milosavljevic}
\author{Amanda L. Simpson}
\author{Valerie A. Smetanka}
\author{Colin P. Feeney}
\author{Shay M. Seguin}
\author{Sieu D. Ha}
\author{Wonill Ha}
\author{Matthew D. Reed}
\affiliation{HRL Laboratories, LLC, 3011 Malibu Canyon Road, Malibu, California 90265, USA}

\begin{abstract}
Precise nanofabrication represents a critical challenge to developing semiconductor quantum-dot qubits for practical quantum computation.
Here, we design and train a convolutional neural network to interpret in-line scanning electron micrographs and quantify qualitative features affecting device functionality.
The high-throughput strategy is exemplified by optimizing a model lithographic process within a five-dimensional design space and by demonstrating a new approach to address lithographic proximity effects.
The present results emphasize the benefits of machine learning for developing robust processes, shortening development cycles, and enforcing quality control during qubit fabrication. 
\end{abstract}

\maketitle


Semiconductor quantum-dot qubits offer an enticing path towards the realization of scalable quantum information processing systems.\cite{Ladd.2017}
Variants based on tensile-strained isotopically pure silicon (800 ppm $^{29}$Si) quantum wells feature modest device footprints ($\approx$0.1 $\mu$m$^2$/qubit), long qubit lifetimes ($>$100 ms)\cite{Dzurak.2014,Vandersypen.2014,Hunter.20150e}, long qdubit dephasing times ($>$2 $\mu$s),\cite{Hunter.20150e} fast control ($<$10 ns)\cite{Andrews.2019}, and rapidly improving two-qubit gate fidelities \cite{Zajac.2018,Watson.2018}. 
Despite this appeal, an inherent difficulty currently limiting progress is meeting the tight fabrication tolerances required by the technology.\cite{Tahan.2020,Sze.1981}
\footnote{The relatively large effective mass of electrons in Si ($m_t* = 0.19m_e$ versus, e.g., $m^* = 0.067m_e$ in GaAs) results in weaker tunneling amplitudes, requiring, in turn, closer quantum dot separations ($\approx$ 150 nm vs. 300 nm in GaAs). $m_e$ is the electron rest mass in vacuum.}
Uncontrolled fabrication processes can impact qubit functionality and performance, leading to both potentially egregious failures (e.g., shorts or opens) as well as more subtle effects (e.g., poorly controlled exchange coupling\cite{Hunter.2015} or unreliable electron loading\cite{Thorbeck.2015}).



\begin{figure}[!tb]
	\includegraphics[width=\columnwidth]{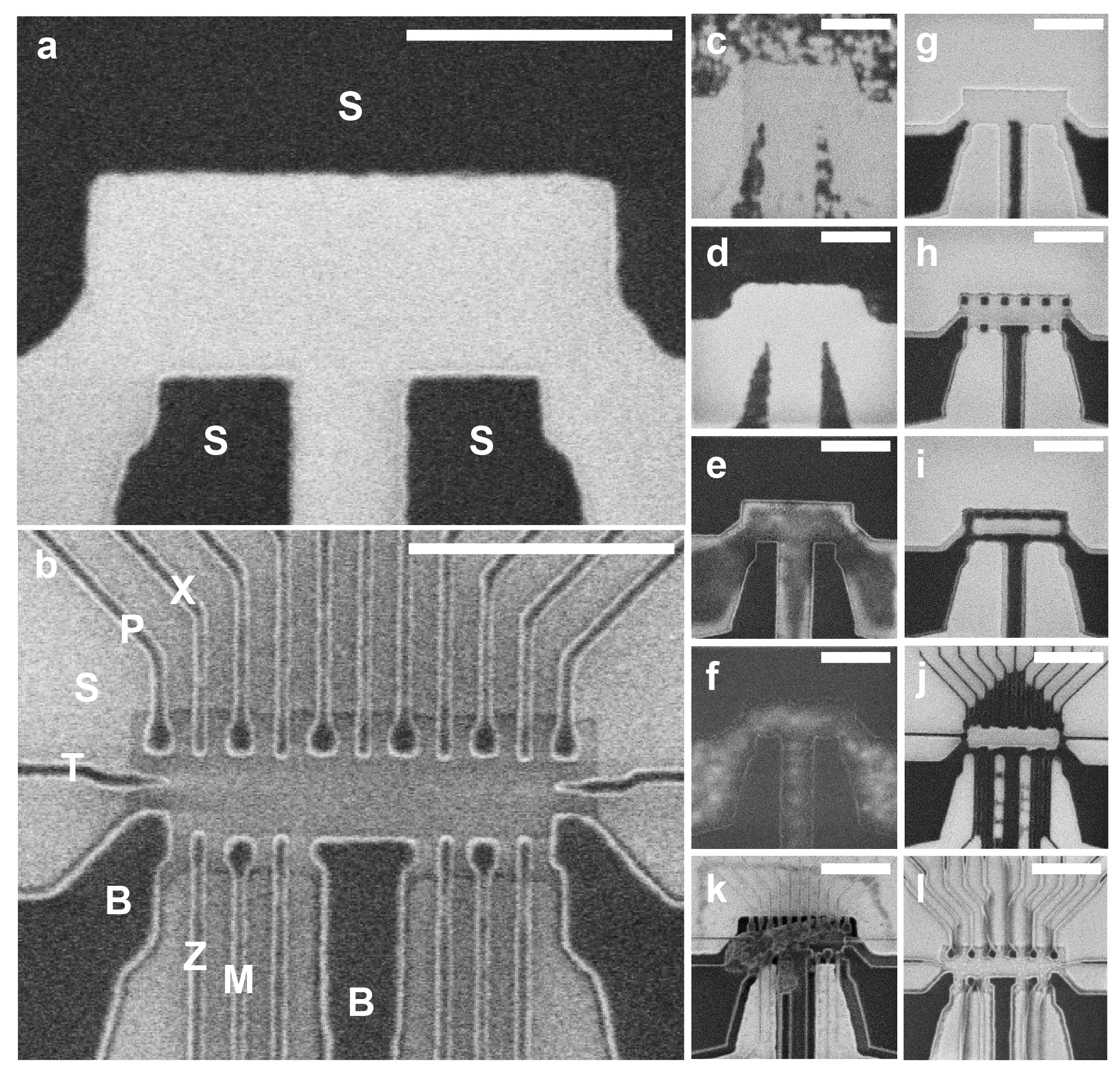}
	\caption{
		\label{fig:SEM} 
		\textbf{Challenges encountered in the fabrication of quantum-dot qubits.} 
		(a) and (b) SEM micrographs of reference devices with well-defined screening S, plunger P, exchange X, tunnel T, sensor Z, measure M, and bath B gates.
		The images in panels (a) and (b) are acquired after pattering screening and quantum-dot gates, respectively. 
		(c)-(l) Critical failures encountered during fabrication include (c-d) underexposed S gate, (e-f) overexposed S gate, (g) missing X and P gates, (h) missing X gates, (i) merged X and P gates, (j) merged leads, (k) particulate contamination, and (l) collapsed patterns. 
		Scale bars measure 500 nm.
	}
\end{figure}

In this Letter, we present a strategy based on machine learning analysis of scanning electron microscopy (SEM) images to optimize processes and enforce quality control during the fabrication of enhancement-mode quantum-dot qubits on Si\textsubscript{1-x}Ge\textsubscript{x} heteroepitaxial foundations.
The approach is showcased using devices comprised of a linear array of six quantum dots and which feature a screening gate architecture\cite{Petta.2015} to confine electrons in buried silicon quantum wells.\cite{Schaffler.1997} 
Quantum information is encoded in superpositions of electron spin states, which are manipulated\cite{Gossard.2005} through pairwise interactions involving exchange gates.\cite{Hunter.2015,Andrews.2019} 
Qubit state readout is performed with the aid of Coulomb blockade and spin-to-charge conversion.\cite{Elzerman.2004,Blumoff.2020} 

Device fabrication involves subtractively defining gates and leads using negative-tone electron-beam lithography. 
After depositing gate materials, hydrogen silsesquioxane (HSQ) films are spun on with thicknesses $20 < d$\textsubscript{HSQ}$< 80$ nm and immediately capped with aluminum layers with thicknesses $d$\textsubscript{Al} between 0 (no cap) and $20$ nm. 
Wafers are then stored in vacuum for a controlled time $t$\textsubscript{HSQ} spanning $10$ to $120$ hr before being exposed in an $100$-keV Raith EBPG5200 e-beam writer with doses $0.4 < D < 2.8$ mC/cm$^2$. 
Development is carried out in MF312 (i.e., tetramethylammonium hydroxide diluted to 4.9\% in water) for $2 < t$\textsubscript{MF312} $< 3$ min.
This process is performed twice: first to define the screening gate and second to pattern the quantum-dot gates and leads.

The lithography are inspected after development using secondary-electron SEM to identify nominally successful outcomes as well as various failure modes.
\footnote{Micrographs are collected using a Hitachi S9380 scanning electron microscope equipped with a Schottky emitter. The primary electron beam is operated at an accelerating voltage of 500 V and a beam current of 7 pA.}
Figures \ref{fig:SEM}(a) and \ref{fig:SEM}(b) show representative examples of well-formed devices after patterning screening and quantum-dot gates, respectively.
Defective devices with malformed screening gates are shown in Figs. \ref{fig:SEM}(c)-(f), with the latter (former) set of panels corresponding to underexposed (overexposed) patterns. 
Figures \ref{fig:SEM}(g)-(i) depict malformed quantum-dot gate morphologies, ranging from entirely absent to completely merged. 
An example of a device with overexposed leads is presented in Fig. \ref{fig:SEM}(j). 
Devices lost to particulate contamination and collapsed resist structures are shown in Figs. \ref{fig:SEM}(k) and \ref{fig:SEM}(l), respectively. 
The qualitative and nonexhaustive failure modes illustrated by Figs. \ref{fig:SEM}(c)-(l) pose challenges to improving qubit yield since they are difficult to address with traditional metrology. 


\begin{figure}[!tb]
	\includegraphics[width=\columnwidth]{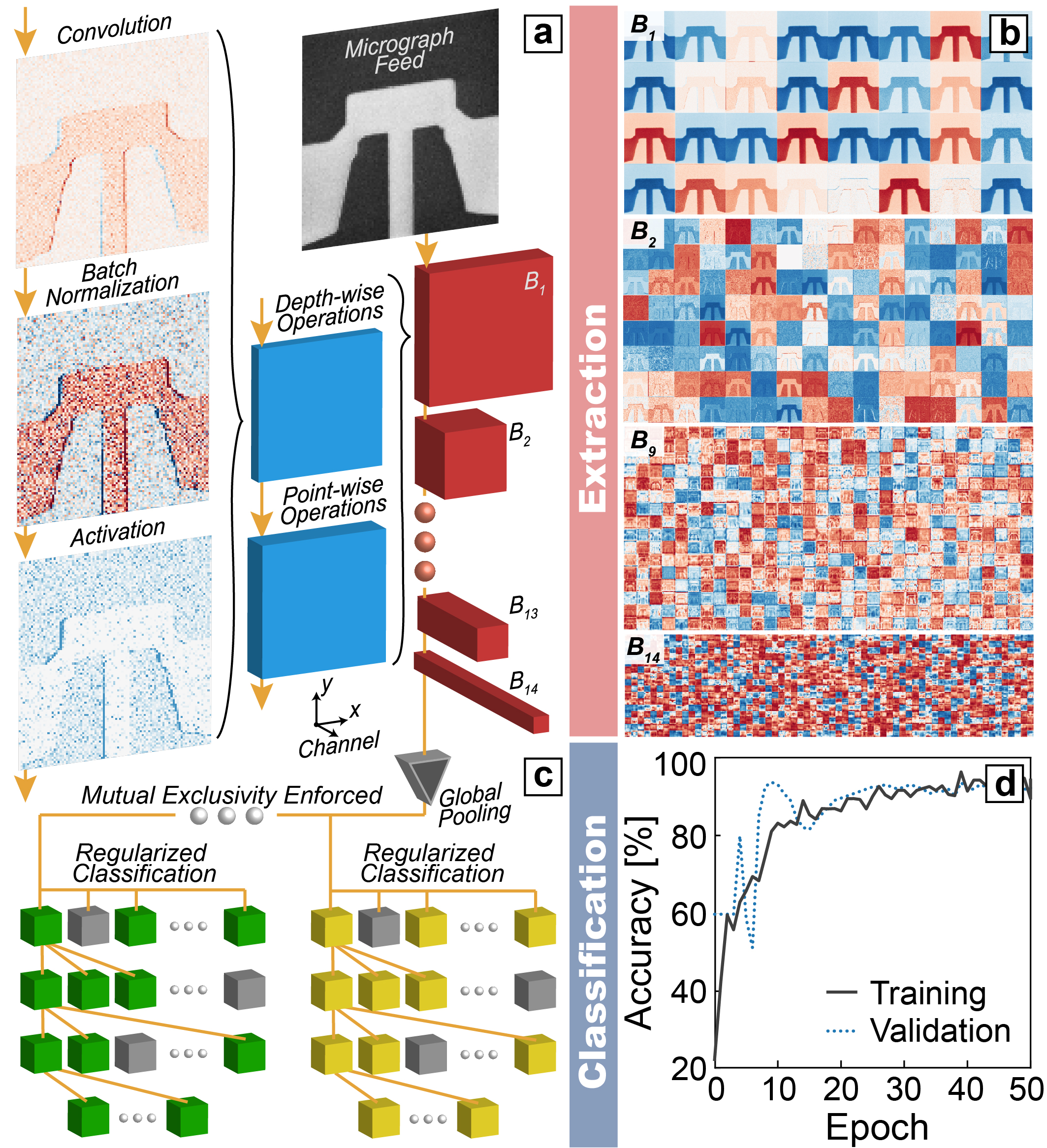}
	\caption{
		\label{fig:CNN} 
		\textbf{Leveraging machine learning in quantum-dot qubit fabrication for root-cause analysis, process optimization, and quality control.} 
		SEM micrographs collected in-line are analyzed using a convolutional neural network consisting of feature extraction and classification subnetworks. 
		(a) Features extraction occurs across fourteen blocks, each levering depth-wise separable operations with independent convolution, batch normalization, and activation layers. 
		Outputs generated by each spatial (depth-wise) layer in block $B_1$ are shown as examples. 
		Similar outputs produced by spatial convolutions in blocks $B_1$, $B_2$, $B_9$, and $B_{14}$ are provided in (b). 
		(c) Classification is carried out on mutually exclusive sub-classes using on regularized three-layer-deep densely-connected networks. 
		(d) Training and validation accuracy converge above 90\% for training epochs $\gtrsim$ 30.
	}
\end{figure}

Our approach to optimizing and monitoring quantum-dot qubit devices in semi-production environments where individual wafers contain thousands of devices has been to design a convolutional neural network which is compatible with high-throughput image interpretation and classification. 
We formulate\cite{Zhang.2016} the network as a two-part system comprised of feature extraction and classification. 

Feature extraction (Fig. \ref{fig:CNN}(a)) is based on a depth-wise separable network architecture\cite{Adam.2017} in which spatial and spectral operations are factored to achieve high classification accuracy with modest computational resources.\cite{Chen.2018} 
The network is assembled as a fourteen-block structure ($B_1$ through $B_{14}$), each with factored operations involving independent convolutions, batch normalization,\cite{Szegedy.2015} and activation\cite{Le.2017}. 
Learning is transferred in the form of over three million parameters pre-trained using public domain images\cite{Fei-Fei.2009}. 
Example outputs from the set of spatial operations within block $B_1$ are provided in Fig. \ref{fig:CNN}(a) for the reference image shown in Fig. \ref{fig:SEM}(a). 
This particular set of operations highlights right edges within images. 
Other features are emphasized in other channels, as the complete set of $B_1$ depth-wise convolution outputs in Fig. \ref{fig:CNN}(b) demonstrates.
Each extraction block $B_i$ typically doubles channels and halves image resolution such that by $B_{14}$, the output adopts the form of a 8$\times$8$\times$1024-dimensional tensor. 
A consequence of these operations is that information becomes increasingly abstract and distributed across a growing number of channels (see Fig. \ref{fig:CNN}(b)). 
That the final pooled\cite{Yan.2013} 1024-dimensional vectors serves as a robust foundation for classification is confirmed by inspecting a gridded\cite{Volgenant.1987} two-dimensional embedded projection\cite{Maaten.2013} of randomly sampled training images organized by structural similarities. 

Classification is handled by a three-layer-deep 1024-neuron-wide ReLU-activated\cite{Le.2017} densely connected network with over two million adjustable parameters (Fig. \ref{fig:CNN}(c)). 
To control over-fitting, dropout\cite{Salakhutdinov.2012,Salakhutdinov.2014} and $L_1$ regularization\footnote{$L_1$ regularization promotes sparse weights, offering a combination of built-in feature selection and reduced computational cost, versus $L_2$ regularization.} are employed.\cite{Courville.2016,Meidani.2018} 
In addition, mutual exclusivity among classes is enforced via separate but similar classification sub-networks. 
Node weights are optimized by teaching the system to recognize thousands of manually labeled SEM images, including Figs. \ref{fig:SEM}(a)-(l). 
Training is monitored by re-purposing one fifth of the training data set for validation. 
By 30 training epochs, both training and validation accuracy exceed 90\% (Fig. \ref{fig:CNN}(d)), indicating that despite the large number of tunable parameters, regularization successfully preserves predictive power.
The resulting network thus constitutes a robust, high-throughput approach to monitoring device quality.

\begin{figure}
	\includegraphics[width=\columnwidth]{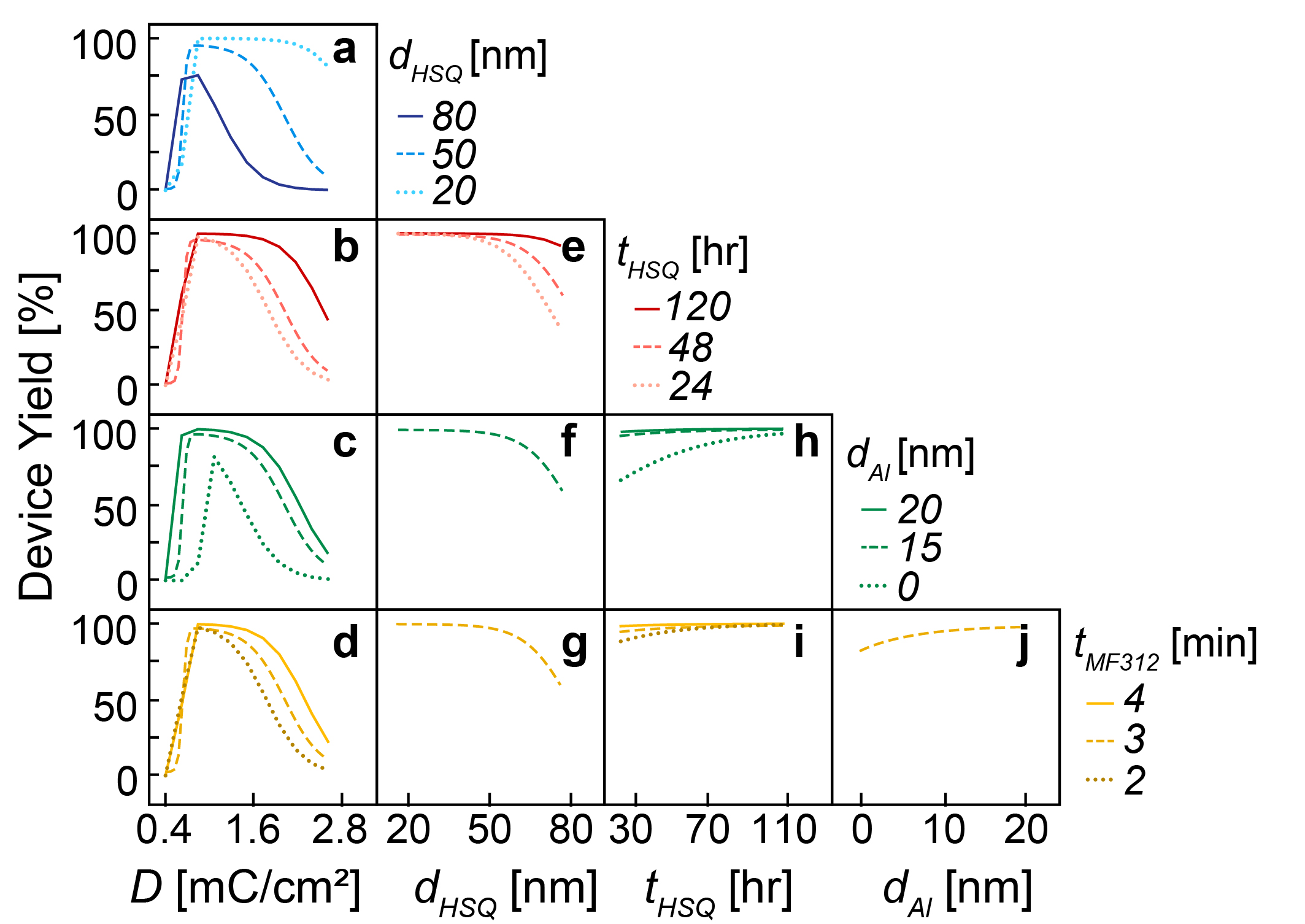}
	\caption{
		\label{fig:Interaction} 
		\textbf{Quantum-dot qubit device yield in a highly interdependent five-dimensional lithographic design space.}
		The panels show percentage of yielding devices $f$ as a function of (a)-(d) exposure dose $D$, (e)-(g) resist thickness  $d$\textsubscript{HSQ}, (h)-(i) exposure delay $t$\textsubscript{HSQ}, and (j) aluminum cap thickness $d$\textsubscript{Al}. 
		The last row of panels comprise sets of curves for which development time $t$\textsubscript{MF312} = 2, 3, and 4 min. 
		For the earlier rows, other experimental factors are varied in accordance with the legend. 
		The lithographic process is maximally stable for $0.5 \lesssim D \lesssim 1.3$ mC/cm$^2$, 
		$d$\textsubscript{HSQ} $\lesssim 50$ nm,
		$t$\textsubscript{HSQ} $\gtrsim 48$ hr, 
		$d$\textsubscript{Al} $\gtrsim 20$ nm, and 
		$t$\textsubscript{MF312} $\gtrsim 3$ min.
	}
\end{figure}


The automated classification capability provided by the neural network affords opportunities to understand and optimize processes in high-dimensional design spaces. 
We demonstrate this by investigating a model process involving lithography with HSQ. 
While HSQ has been successfully employed as an etch-resistant negative-tone resist to pattern nanostructures with $\lesssim10$ nm resolution,\cite{Kurz.2003,Hagen.2007,Rooks.2009,Chong.2011} it suffers from a number of outstanding challenges. 
Chief among these is an extreme sensitivity to exposure and development conditions.\cite{Delft.2002,Krchnavek.2006,Cui.2006} 

To characterize HSQ and establish a robust lithography process suitable for quantum-dot qubit fabrication, we explore a five-dimensional design space spanned by exposure dose $D$, HSQ film thickness $d$\textsubscript{HSQ}, aluminum cap thickness $d$\textsubscript{Al}, exposure delay $t$\textsubscript{HSQ}, and development time $t$\textsubscript{MF312}.
Lithographic quality, as determined via our neural network following screening gate patterning, is correlated with experimental variables using a logistical regression.
The analysis models over five thousand unique design points to identify, with an accuracy of 92\%, a response surface which optimally describes device yield as a function of the predictors.
Yield curves obtained by interpolating one-dimensional sections through the five-dimensional manifold are plotted in Figs. \ref{fig:Interaction}(a)-(j).
\footnote{Curves are omitted from panels for which interactions between variables were not investigated.}
The dashed line in each panel represent reference yield curves centered about the design point $D$ = 1.0 mC/cm$^{-2}$, $d$\textsubscript{HSQ} = 50 nm, $d$\textsubscript{Al} = 15 nm, $t$\textsubscript{HSQ} = 48 hrs, and $t$\textsubscript{MF312} = 2 min. 
The additional solid and dotted lines are yield curves where one of the predictors has been skewed according to the figure legend.


The reference profile in Fig. \ref{fig:Interaction}(a) shows the evolution of network-classified lithography yield $f$ as a function of exposure dose $D$. 
The percentage of correctly exposed devices increases rapidly from zero to a peak of $f \approx 75\%$ at $D \approx$ 1.0 mC/cm$^2$, before decreasing asymptotically back to zero. 
The lower threshold dose $D_l$, defined as the dose above which device yield exceeds 50\%, is found to be $\approx$ $0.6$ mC/cm$^2$, in agreement with values assessed previously from profilometry measurements.\cite{Rooks.2009,Ohkubo.2011,Chong.2011} 
The upper dose limit, defined analogously, is $2.0$ mC/cm$^2$, yielding a dose exposure window $\Delta D \equiv D_u - D_l$ = 1.4 mC/cm$^2$. 

Also plotted in Fig. \ref{fig:Interaction}(a) are yield curves corresponding to resist thickness values $d$\textsubscript{HSQ} $=$ 20 and 80 nm. 
The three curves combined indicate that not only does increasing the resist thickness monotonically reduce both lower $D_l$ and upper $D_u$ exposure dose limits, but $D_u$ decreases faster than $D_l$, leading to a restricted span of dose values over which properly defined patterns are obtained (i.e., a narrower dose latitude $\Delta D$). 
This unorthodox behavior, which corresponds to a thickness-dependent exposure sensitivity, stems from a combination of factors including self-limiting development processes\cite{Rooks.2009,Kim.2009} and collateral exposure from secondary electrons generated within the resist\cite{Chong.2011}. 
Since spun-on resist preferentially accumulates in trenches and tapers near step edges,\cite{Kuo.2011} thickness-induced sensitization can critically impact lithographic yield on uneven surfaces.
In extreme cases, patterns in trenches merge before features on mesas are defined (e.g., Fig. \ref{fig:SEM}(i)). 


HSQ is an intrinsically unstable compound.\cite{Grove.2005} 
The instability stems from highly mobile and reactive hydrogen species, which reside on the vertices of the cuboid silsesquioxane structure.\cite{Toskey.1998} 
Exposure to energy in the form of heat, electrons, or ultra-violet photons triggers the release of hydrogen as radicals. 
The radicals diffuse catalyzing redistribution reactions which cross-link neighboring molecules into base-insoluble hydrogen-enriched SiO\textsubscript{x} oligomers.\cite{Leone.2006,Drift.2009,Schuck.2010} 
Ambient conditions are insufficiently inert to completely impede the reactions,\cite{Grove.2005} 
resulting in limited shelf life\cite{Delft.2002} and transient exposure behaviors\cite{Krchnavek.2006}.

Figure \ref{fig:Interaction}(b) shows the impact delaying exposure by $t$\textsubscript{HSQ} = $24$, $48$, and $120$ hrs has on the percentage of well-formed devices $f$ for doses $D$ between $0.4$ and $2.8$ mC/cm$^2$. 
The upper dose limit above which half of the patterns are overexposed increases with $t$\textsubscript{HSQ} from $D_u$ = $1.8$ ($24$) to $2.5$ mC/cm$^2$ ($t$\textsubscript{HSQ} = $120$ hr) as an increasing fraction of neighboring features begin bridging together (e.g., Figs. \ref{fig:SEM}(e), \ref{fig:SEM}(f), \ref{fig:SEM}(i), and \ref{fig:SEM}(j)). 
Improved device yield at longer $t$\textsubscript{HSQ} is consistent with published reports of lithographic contrast being higher for HSQ films for which exposure was delayed.\cite{Kurz.2003,Krchnavek.2006}

At a near-optimal exposure dose $D = 1.0$ mC/cm$^2$, device yield is lowest and most susceptible to varying with time when the resist is thick ($d$\textsubscript{HSQ} $\gtrsim$ $80$ nm, Fig. \ref{fig:Interaction}(e)) and uncapped ($d$\textsubscript{Al} $= 0$ nm, Fig. \ref{fig:Interaction}(h)). 
The former observation is consistent with the cumulative and diffusive nature of hydrogen-based catalysis. 
The latter finding highlights the effectiveness of aluminum barriers in allowing the film to stabilize in an equilibrated state through the suppression of hydrogen out diffusion.\cite{Chua.2000,Alford.2000} 

 
Collectively, Figures \ref{fig:Interaction}(a)-(j) establish that device yield is maximized for $0.5 \lesssim D \lesssim 1.3$ mC/cm$^2$, $d$\textsubscript{HSQ} $\lesssim 50$ nm, $t$\textsubscript{HSQ} $\gtrsim 48$ hr, $d$\textsubscript{Al} $\gtrsim 20$ nm, and $t$\textsubscript{MF312} $\gtrsim 3$ min. 
Aluminum capping films not only boost maximum yield from $f \approx 75\%$ at $d$\textsubscript{Al} = $0$ nm (no cap) to $f = 100\%$ with for $d$\textsubscript{Al} = 20 nm, but also widen the process window from $\Delta D \approx 0.5$ to $1.5$ mC/cm$^2$ over the same range (see Fig. \ref{fig:Interaction}(c)). 
Aggressive development utilizing concentrated hydroxides and longer processing times also extend the upper dose bounds $D_u$ by dissolving collaterally exposed resist (see Fig. \ref{fig:Interaction}(d)).


\begin{figure}
	\includegraphics[width=\columnwidth]{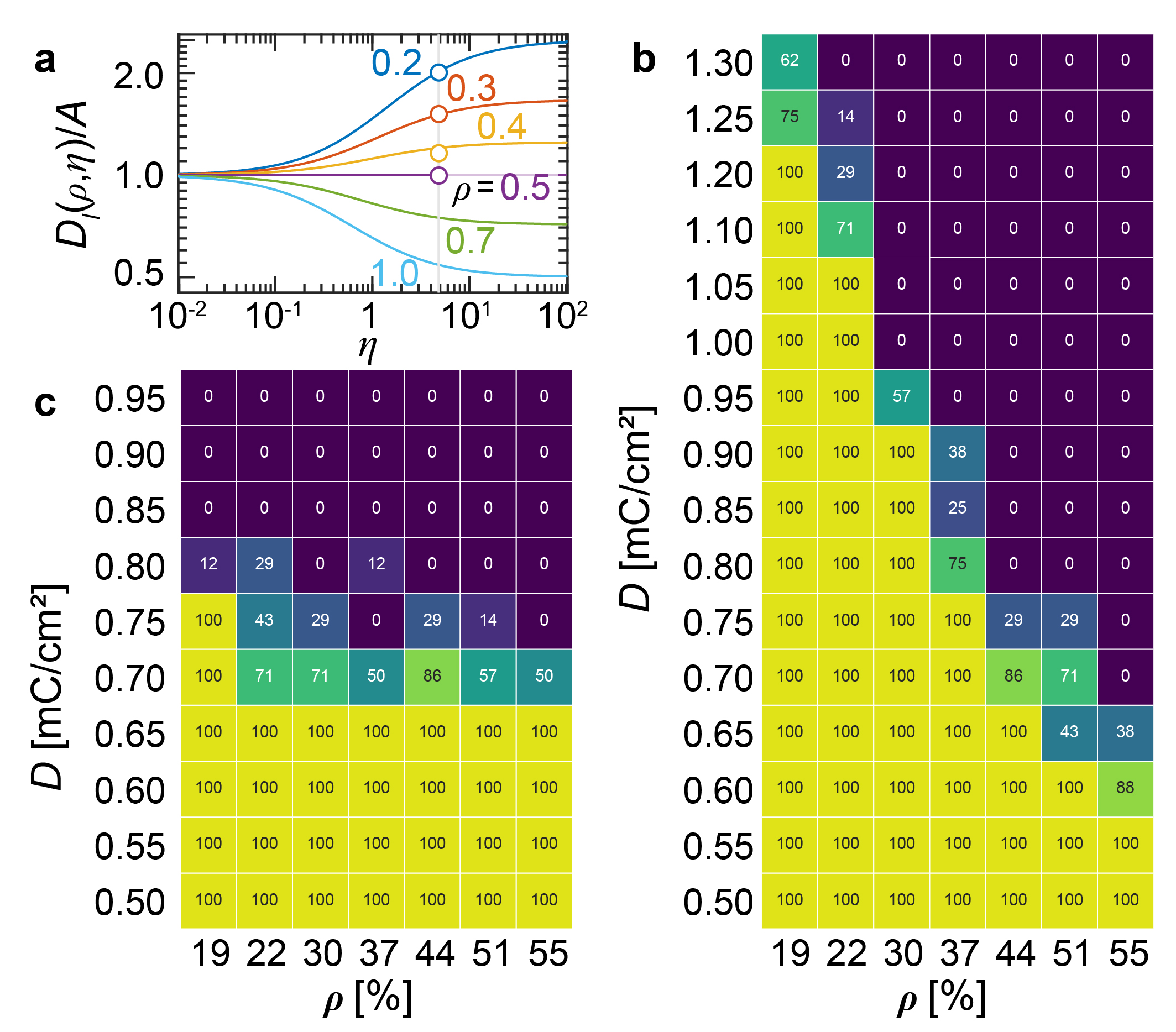}
	\caption{
		\label{fig:PEC} 
		\textbf{Machine learning approach to proximity effect correction in electron-beam lithography.} 
		(a) Model describing the change in onset exposure dose $D_l$ with local pattern density $\rho$ and back-to-forward electron scattering energy ratio $\eta$. 
		Percentage of underexposed devices as a function of exposure dose $D$ and filling fraction $\rho$ (b) before and (c) after accounting for proximity effects.
	}
\end{figure}

Even with optimized exposure and development procedures, a complication that arises when carrying out electron-beam lithography is that regions with sparse features will generally require higher exposure doses than areas where features are denser (e.g., Fig.\ref{fig:SEM}(j)).\cite{Chang.1975} 
Conventionally, the effect stems from an energy point spread function (PSF) which extends many beam widths away from the exposure site. 
Approaches to address proximity effects to date have relied predominately on measuring linewidths of dedicated monitor structures.\cite{Murai.1992,Hofmann.2018} 
We now demonstrate an alternative approach that leverages our machine learning algorithm to identify onset exposure doses as a function of local pattern densities naturally occurring in the leads of our devices. 
\footnote{Leads in our devices are spaced at a pitch of $\sim$ 70 nm and exhibit widths which vary between $\sim$ 10--40 nm.}

The evolution of onset exposure dose $D_l$ with local pattern density $\rho$ is traditionally described by a series-expanded Gaussian convolution model,\cite{Hofmann.2014}
\begin{equation}
   \label{eq:PEC}
   D_l(\rho,\eta) = A \frac{1 + \eta}{ 1 + 2 \rho \eta },
\end{equation}
in which $\eta$ is the energy ratio of backscattered and forward scattered electrons and $A$ sets the baseline dose at half filling, i.e. $D_l(\rho=50\%)$. 
A graphical representation of Eq. \ref{eq:PEC} is provided in Fig. \ref{fig:PEC}(a) as constant-$\rho$ plots of $D_l(\rho,\eta)/A$ versus $\eta$. 
For $\eta \lesssim$ 0.1, all curves are approximately equal to unity, indicating that threshold doses are independent of pattern density. 
As $\eta$ increases between $0.1 \lesssim \eta \lesssim 10$, the curves fan out with $D_l$ increasing (decreasing) for sparse (dense) patterns. 
In the limit of arbitrarily large $\eta$, $D_l$ approaches $A/(2\rho)$ asymptotically. 

Figure \ref{fig:PEC}(b) shows $f_u$ the percentage of underexposed devices as a function of dose $D$ and local pattern density $\rho$. 
Experimental onset doses $D_l$, defined as the highest value for which $f_u < 50\%$, decay rapidly with $\rho$ from $1.30$ mC/cm$^2$ at $\rho = 0.19$ to $0.60$ mC/cm$^2$ for $\rho = 0.55$. 
$\eta$, which controls the rate of decay, is determined graphically from Fig. \ref{fig:PEC}(a) and rigorously from robust nonlinear regression analyses ($R^2$ $\gtrsim$ 0.99) to be $\eta \simeq 5$. 
For comparison, Monte Carlo simulations of electron-sample interactions yield $\eta \approx 0.1$;\cite{GenISys.2020} for 500-nm-thick ZEP
\footnote{ZEP is a common positive-tone electron-beam lithography resist based on methyl styrene and chloromethyl acrylate copolymer.} 
and polymethyl methacrylate, $\eta$ ranges between 0.6 and 1.1, depending on the substrate material.\cite{Brown} 
The significantly larger value of $\eta$ observed here reflects the complex nature of proximity effects in HSQ, which involve not only the diffusion and scattering of electrons, but also hydrogen radicals, which modulate cross-linking energy thresholds.\cite{Ohkubo.2011}

A plot of underexposed device percentages $f_u(D,\rho)$ after accounting for proximity effects is provided in Figure \ref{fig:PEC}(c). 
The results are obtained by locally modulating base doses within patterns with a spatially varying dose multiplier that depends on the output of convolutions between computer-aided designs and an $\eta$-corrected energy PSF. 
In distinct contrast to the results obtained for uncorrected exposures (i.e., Fig. \ref{fig:PEC}(b)), the revised procedure produces threshold doses that are constant at $D_l \approx 0.7$ mC/cm$^2$ and independent of pattern density $\rho$, thus enabling a span of dissimilar geometries to be patterned concurrently.


In summary, we have designed and trained a convolutional neural network to quantify qualitative failure modes impacting quantum-dot qubit fabrication yield. 
The algorithm has been leveraged to understand and optimize a highly-correlated five-dimensional lithographic process space. 
A novel strategy for addressing electron beam proximity effects has also been successfully demonstrated. 
The present results highlight the effectiveness of machine learning in guiding root-cause analysis, process development, and quality control of complex quantum devices.

\section{Acknowledgments}

We thank John R. Lowell, Reed W. Andrews, and Matthew G. Borselli for valuable discussions. We gratefully acknowledge technical support of HRL Laboratories Microfabrication Technology Laboratory engineers, especially Jack A. Crowell, Larken E. Cumberland, Chuong V. Dao, Joshua M. Doria, Elias A. Flores, Vinh S. Ho, Rosalinda M. Ring, Golam Sabbir, Reggie D. Salinas, Mariano J. Taboada. This research was carried out with financial support from The Boeing Company.

\bibliographystyle{naturemag}
\bibliography{library}

\end{document}